\documentclass[USenglish]{article}

\usepackage[utf8]{inputenc}%(only for the pdftex engine)
\usepackage[small]{dgruyter}
\usepackage{microtype}
\usepackage{subfig}

\begin{document}

  \articletype{...}

  \author*[1]{Karoline Busse}
  \author[2]{Sabrina Amft}
  \author[1]{Daniel Hecker}
  \author[3]{Emanuel von Zezschwitz} 
  \runningauthor{Busse, Amft, Hecker, von Zezschwitz}
  \affil[1]{University of Bonn}
  \affil[2]{Leibniz University Hannover}
  \affil[3]{University of Bonn \& Fraunhofer FKIE}
  \title{``Get a Free Item Pack with Every Activation!''}
  \runningtitle{Do Incentives Increase the Adoption Rates of Two-Factor Authentication?}
  \subtitle{Do Incentives Increase the Adoption Rates of Two-Factor Authentication?}
\abstract{Account security is an ongoing issue in practice. Two-Factor Authentication (2FA) is a mechanism which could help mitigate this problem, however adoption is not very high in most domains. Online gaming has adopted an interesting approach to drive adoption: Games offer small rewards such as visual modifications to the player's avatar's appearance, if players utilize 2FA.
In this paper, we evaluate the effectiveness of these incentives and investigate how they can be applied to non-gaming contexts.
We conducted two surveys, one recruiting gamers and one recruiting from a general population. In addition, we conducted three focus group interviews to evaluate various incentive designs for both, the gaming context and the non-gaming context.
We found that visual modifications, which are the most popular type of gaming-related incentives, are not as popular in non-gaming contexts. However, our design explorations indicate that well-chosen incentives have the potential to lead to more users adopting 2FA, even outside of the gaming context.}

  \communicated{...}
  \dedication{...}
  \received{...}
  \accepted{...}
  \journalname{...}
  \journalyear{...}
  \journalvolume{..}
  \journalissue{..}
  \startpage{1}
  \aop
  \DOI{...}

\maketitle

\section{Introduction}
\begin{figure}[t!] 
    \centering 
    \subfloat[]{ 
        \includegraphics[width=0.5\textwidth]{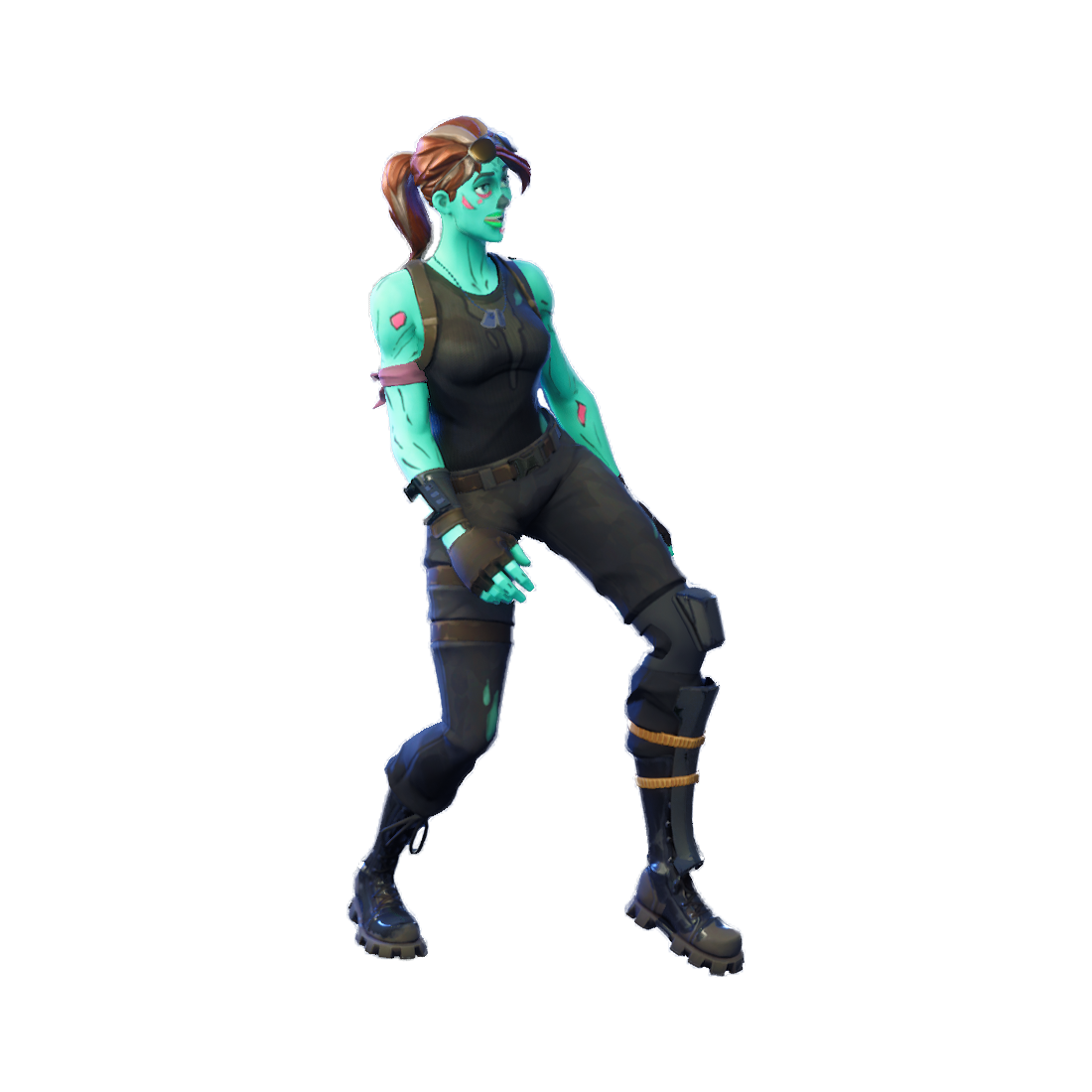}}
    \subfloat[]{ 
        \includegraphics[width=0.5\textwidth]{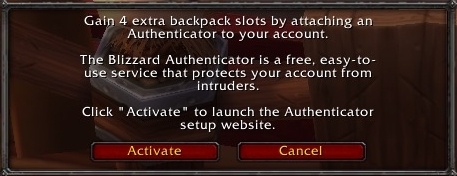}} 
    \caption{Examples of incentives for adopting 2FA. (a) shows an exclusive dance emote for player characters in \textit{Fortnite}~\protect\cite{fortnite}, while (b) shows a promoting message for a small gameplay advantage, namely more item space in \textit{World of Warcraft}~\protect\cite{wow}. Images used with permission, see Acknowledgement for full copyright statements.}
    \label{fig:examples}
\end{figure}

The most widespread way to authenticate users is to require them to input a combination of an identifier such as a e-mail-address or a username as well as a password.
Often, these passwords are either very easy to guess~\cite{adams1997making, kelley2012guess} or can be disclosed through several other means, e.g. phishing, data leaks or insecure storage~\cite{mirante2013understanding, walters2014cyber}. 
Two-factor authentication (2FA) is a security mechanism that was designed to help increase security~\cite{joram2009multiple} by adding a second factor to the authentication instead of relying on secret knowledge, such as a passwords, as a single factor. Common manifestations of this second factor include SMS and e-mail notifications, dedicated smartphone applications, or hardware tokens.

When research looks into the process of adopting 2FA, usually the transition process and usability issues within larger organizations are accompanied, with a focus how users react and adopt to the change~\cite{colnago2018its, libicki2011influences, weidman2017like}. %1) perceived as secure but annoying 2)  
From these papers we know that 2FA is perceived as secure but annoying, and that forced 2FA adoption in the workplace sometimes leads to increase usage rates for personal accounts~\cite{colnago2018its}.
Less research has been done for individuals who choose whether to adopt 2FA in their private life~\cite{krol2015they}.

In the past there have been attempts to increase the 2FA adoption rates by offering rewards to users who choose to do so. This is especially widespread in the gaming sector and only very rarely found with other companies. Possible incentives include e.g. visual benefits such as virtual pets or gameplay advantages such as premium currency or access to special ingame vendors (see Figure~\ref{fig:examples} for examples).
For most of these incentives, two rules apply: 1) They can only be received by complying and adopting 2FA, and 2) if the player chooses to deactivate 2FA again, the rewards are withdrawn from their access.
We argue that there are no obvious reasons why other sectors did not yet adopt incentives in the same way the gaming business did.
One hypothesis might be that users who play video games might tend to be more interested in IT and security mechanisms and are therefore faster to comply and activate 2FA. 
Another reason could be that the accounts in this sector often do not only contain monetary value as players bought games, premium currency, or items; but also that players might have spent hundreds of hours to establish a rank or a well-working game status that is often very desired by criminals. 
In fact, there is a huge market for valuable accounts as just buying a high-leveled status saves other users much time. 
Due to this, video game accounts are often targeted by malicious actors~\cite{marrujo2018fraud, trend2015data}.

Typical incentives in video games are cosmetic modifications like redesigns for characters or items~\cite{wow,fortnite,rainbowsixsiege}, cosmetic companions~\cite{gw2,wow}, gameplay advantages like larger inventories~\cite{wow}, in-game premium currencies~\cite{swtor}, multipliers on certain factors such as experience gains~\cite{wildstar}, or access to special items or in-game vendors~\cite{swtor} (cf. Figure~\ref{fig:examples}).
One notable exception is Valve's store and community platform Steam.
Without the use of 2FA, item trades or sales between players are on hold for up to fifteen days before a trade is concluded~\cite{steam}.
Similarly, Electronic Arts requires users to enable 2FA for their accounts before granting access to the online or mobile versions of FIFA Ultimate Team since 2015~\cite{fut}.

Since these kinds of incentive mechanics are almost exclusively found in the gaming sector, we assume that we can learn from their presumed success.
Therefore, we chose to investigate this topic further and posed the research question:

\begin{quote}
How can we transfer (successful) incentives for adopting two-factor authentication from gaming to non-gaming contexts?    
\end{quote}

During the course of this study, we reached out to developer studios and publishers of video games that employ incentives for 2FA adoption.
Our goal was to get data on adoption rates before and after the introduction of an incentive.
In addition, we planned a short interview on 2FA and their experiences with it.
Sadly, no studio answered our request, and personal acquisition at the Gamescom 2018 convention did also not lead to any responses.

In response to this lack of data from the service provider side, we put strong emphasis on user research.
We conducted two survey studies with a gaming-focused ($N=462$) as well as a general population sample ($N=288$) about 2FA and incentives for adopting it.
From these surveys, we extracted design proposals for transferring incentive models from gaming to non-gaming contexts and tested these in a focus group study with three groups and a total of 15 participants.

We found that incentives increase the adoption rates for 2FA for services that employ them, but users rather self-report that they activated 2FA for security reasons and not for the incentives.
The most often encountered type of incentive in gaming contexts, namely cosmetic modifications or items, was perceived least attractive in our focus group study, suggesting that such approaches are not directly applicable to non-gaming contexts.
In contrast, small monetary or service-focused incentives were considered most attractive in a non-gaming scenario.
Through discussion and comments, we identified a security-privacy trade-off in users' mental models when it comes to adapting 2FA, suggesting that offering users a selection of 2FA methods along with basic educational material would lead to higher adoption rates.

\section{Related Work}

\subsection{Two-Factor Authentication}

While there is not much research about the combination of 2FA with incentives to adopt it, there are several studies concerning adoption rates, effectiveness and usability.

In 2014, Gunson et al. conducted a study that compared single-factor authentication (1FA) to two-factor authentication (2FA) for automated banking purposes. 
The 1FA mechanism required users to recall secret knowledge, a few digit PIN, on the telephone. 
This is common procedure for telephone banking processes. 
The 2FA group however was additionally tasked to enter a code that was transmitted through a hardware token. 
The researchers found that while 2FA was perceived more secure, it was also reported to be less usable and convenient when compared to the 1FA mechanism as it took longer and required a bit more work~\cite{gunson2011user}.

Cristofaro et al. made a more general survey in 2015, trying to compare different approaches to 2FA in terms of usability. 
They conducted a survey that recruited 219 participants on Amazon MTurk and chose to compare three different kinds of 2FA: hardware security tokens, codes send via e-mail or SMS, and apps like Google Authenticator. 
While all 2FA mechanisms were overall perceived as usable, most users did only adopt it because they were forced to do so (37-44\% depending on the 2FA solution), while 35-53 \% were using it voluntarily.
Only 9-19\% responded that they use 2FA due to incentives~\cite{cristofaro2013comparative}.

In 2018, Colnago et al. released a paper on how the \textit{Duo} 2FA system was distributed at their university in Pennsylvania.
While students had the option to adopt the system, it was made mandatory for all of the universities employees. 
They conducted two surveys, one before and one after the mandatory enrollment of the service, finding that most users perceived the system to be a bit annoying, but still easy to use. 
It also became visible that after using Duo for a while, users became accustomed to 2FA and sometimes even started using it on other accounts as well~\cite{colnago2018its}.

From this work, we can learn that 2FA poses an increase in cognitive load during the authentication process.
This disadvantage in perception can be offset in some cases by the increase in perceived security.

\subsection{Gamification}

Offering in-ecosystem incentives can be considered a form of \textit{gamification}.

The concept of gamification has been around since the 1980, but in 2011, Deterding et al. analytically defined the term as
\begin{quote}
    The use of design elements characteristic for games in non-game contexts.~\cite{deterding2011from}
\end{quote}
The approach of offering incentives can be considered a design element which is characteristic for games, such as a quest reward.
However, the contexts in which incentives for 2FA activation are already employed are all gaming-related, so this work aims at developing gamification approaches which incentivize 2FA activation outside of the gaming context.

Gamification in security has been mostly done for training and education purposes.

Francia et al. for example developed two games about password strength and phishing awareness.
They used visual metaphors like walls made from different materials for symbolizing password strength.
The two games were self-explanatory and required only a short period of learning~\cite{francia2014gamification}.
Additional work on educational games on phishing includes \textit{Anti-Phishing Phil} by Sheng et al.~\cite{Sheng2007anti}, and \textit{What.Hack} by Wen et al.~\cite{Wen2017what}.

Dabrowski et al. evaluated their own approach of enhancing university security education with gamification elements, namely leaderboards and user ranks with additional incentive for achieving the highest rank.
The authors did both short-term and long-term evaluations of the system using surveys during and after the course.
Results show that the gamification elements motivate the course participants to intensively engage with the course material and raise general interest in security.
However, some of these effects were also attributed to the stereotype of ``hacking'' as something adventurous, as portrayed in the media~\cite{dabrowski2015leveraging}.

\subsection{Increasing User Motivation With Incentives}

It is possible to use virtual rewards such as badges to motivate users, as Anderson et al. showed in 2013.
In their study they used a website similar to Stack Overflow, generated several tasks that included different kinds of participation with the community and awarded badges for users that took part in it. 
Their results show that not only did the badges increase the users' motivation and participation rates, they were also able to predict to a certain degree what a user would do~\cite{anderson2013steering}.

Barata et al. conducted a study in 2013 where they added gamification elements to a master degree university course in engineering, including leaderboards, scores and levels. 
When compared to the same course in the previous year and other university courses, results showed that students and teachers seemed more content with the course and their achievements. 
Several other statistics such as attendance or preparation for courses also increased~\cite{barata2013engaging}.

These studies show that incentives can be a powerful tool to motivate and steer users, even if they only exist meaningfully within a single platform.
Based on the assumption that this could also be applied to incentives for 2FA that we find in video games, this is the first paper to study the feasibility of using incentives for increasing 2FA adoption for non-gaming services.

\section{Online Surveys}

In order to get a better overview on users' attitudes towards incentives for adopting 2FA in gaming, we developed an online survey that was deployed in various online gaming communities in January and February 2018.
Afterwards, this survey was modified and deployed again on Amazon MTurk to recruit a more general sample in January and February 2019.

\subsection{Methodology}

Based on our research question, we created a 14-question online survey consisting mostly of multiple choice and Likert-scale questions.
We asked participants what services from a pre-selected list they use, whether they have 2FA enabled for these services, and some general perception and usability questions regarding various 2FA methods.
Afterwards, we asked about different types of incentives and how they would influence adoption as well as deactivation of 2FA.

We distributed the survey via social media and through posts in the following subreddits: r/SampleSize, r/Blizzard, r/Steam, r/WoW and r/GuildWars2.
As this survey was conducted as part of a student project of one of our authors, we weren't able to compensate the participants of this study.
After the survey was evaluated, a follow-up post with summarized results was posted in the respective communities as a token of appreciation.

For the general audience, some modifications were made to the survey.
First, a question about the term \textit{valve} was included with different possible meanings provided to test whether or not the person would recognize it as the company behind the game marketplace Steam, followed by a second question that directly asked if the user enjoys playing video games.
These were added in order to be able to better differentiate between users that were similar to the participants of the first study, e.g., having an affinity for video games and maybe already familiar with incentives, or if they were part of a more general population.
We specifically asked about Valve and Steam, which might be best known to PC gamers, but less likely to console or mobile gamers.
This was a conscious decision because most of the games who employ 2FA mechanics are PC-centered or even exclusive, such as World of Warcraft.
In addition, questions and answer options regarding incentives were modified to include not only gaming-related options.
To counter-balance our participant's mental load of this extended survey, we chose to change our Likert-scale questions from 7-point to 5-point, trading a finer resolution with (hopefully) more accurate answers.
The full question set of both surveys can be found in the Appendix.

After a pre-test with 20 personal contacts which led to minor improvements in the survey design, we hosted the modified survey on Amazon Mechanical Turk (MTurk) with the following participant requirements: Participants had to have a HIT approval rate of at least $90\%$ and needed to be based either in Canada, Germany, the USA, or the United Kingdom. We chose the geographic locations for potential participants in accordance with our demographic sample from the gaming-focused survey.
All MTurk participants were compensated with USD 2.00 for their work.

\subsection{Results}
\subsubsection{Gaming Community}
\label{sec:res_gaming}

For our gaming-related sample, we could collect 594 data sets, from which 462 were valid.
The participants were on average $27$ years old ($\sigma=7.2$).
Of all participants, $76\%$ self-identified as male, $19\%$ as female, and $4\%$ as non-binary. 
Most people were living in the United States of America ($36\%$), $14\%$ were from Germany, $7\%$ from Canada, and $7\%$ from the United Kingdom.
Other countries were represented by less than 5\% of participants and are omitted here.

\begin{table}[t]
    \centering
    \begin{tabular}{llrr}
        Service & Users & Users with activated 2FA\\
         \midrule
         Blizzard & $56.7\%$ & $43.1\%$\\
         Discord & $71.4\%$ & $26.0\%$ \\
         Facebook & $50.9\%$ & $24.5\%$\\
         GOG & $25\%$ & $6.9\%$\\
         Guild Wars 2 & $65.6\%$ & $55.0\%$ \\
         Nintendo & $23.4\%$ & $3.9\%$ \\
         Origin & $33.5\%$ & $11.0\%$ \\
         PSN & $17.3\%$ & $5.2\%$ \\
         Reddit & $87.7\%$ & $11.5\%$ \\
         Slack & $16.0\%$ & $4.3\%$ \\
         Steam & $88.1\%$ & $62.6\%$ \\
         Telegram & $13.2\%$ & $5.4\%$ \\
         Twitter & $43.3\%$ & $16.9\%$ \\
         Wargaming & $1.3\%$ & $0.6\%$ \\
         WhatsApp & $36.8\%$ & $6.1\%$ \\
         Xbox Live & $11.0\%$ & $4.1\%$ \\
    \end{tabular}
    \caption{Usage and adoption rates for various gaming-related surveys within a gaming-centered population. Both percentages in relation to all participants, $N=462$.}
    \label{tab:usage_gaming}
\end{table}

At the beginning of the survey, we asked participants to check which services (from a pre-compiled list of 2FA-enabled services) they use and for which of them they have 2FA enabled.
The results are presented in Table~\ref{tab:usage_gaming}.
Only 21 participants stated that they do not use 2FA for any of their accounts.
Overall, gaming-related accounts have a higher adoption rate for 2FA than non-gaming accounts, with those that prominently offer an incentive (Blizzard, Guild Wars 2, Steam) having even higher rates.
The participants had the opportunity to add other services for which they use 2FA.
Twenty-three people named Google services like Google Mail, five mentioned online banking and one person claimed they would use 2FA on all their accounts if the service would support it.
Other online platforms mentioned were Dropbox, GitHub, GitLab, Microsoft services, e-mail provider and games like Star Wars: The Old Republic, Wildstar, EVE Online, and Final Fantasy XIV.

Regarding different 2FA methods, SMS is the most widespread method of 2FA with 311 reported users ($67.3\%$ of all participants). 
Google Authenticator is used by $54\%$ of the participants, while service specific apps are used by $50\%$. 
A portion of $44.8\%$ is using e-mail as a way of getting the second factor. 
Hardware tokens are way less widespread with only $6.5\%$ of participants using them.

Afterwards, participants were asked to rate the different presented 2FA methods regarding their perception of how convenient the usage of those methods is and how secure they regard a method, both on a 7-point Likert scale.

SMS was rated very convenient with an average score of $5.14$ ($\sigma=1.88$, median=6), but received only an average rating of $4.85$ ($\sigma=1.79$, median=5) in perceived security. 
Receiving access codes via mail received an average convenience score of $4.55$ ($\sigma=1.78$, median=5) and an average security rating of $4.11$ ($\sigma=1.55$, median=4).
The Google Authenticator received both a slightly higher convenience rating ($\mu=5.32$, $\sigma=1.70$, median=6) as well as a slightly higher security rating ($\mu=5.72$, $\sigma=1.28$, median=6) on average compared to SMS. %TODO significance check 
Service specific applications were rated on average $4.79$ in convenience ($\sigma=1.99$, median=5) and $5.77$ in security ($\sigma=1.39$, median=6). 
The lowest convenience score ($\mu=3.38$, $\sigma=1.97$, median=3), but the highest perceived security ($\mu=6.17$, $\sigma=1.40$, median=7) was given to hardware tokens.

In the next part of the survey, participants were asked about what motivates them to enable 2FA on their accounts.
The enhancement of the account security was a motivational aspect for $88.1\%$ of participants. 
The second most indicated motivation for using 2FA is a high monetary value attached to the corresponding account for a total of $50.9\%$. The possibility of circumventing restrictions motivates $27.3\%$ of participants, whereas visual bonuses are only able to attract $21.9\%$ of participants.
Just $5\%$ of participants stated that they have activated 2FA to gain gameplay advantages.
Afterwards, we presented hypothetical scenarios that might influence the adoption of 2FA and asked participants how likely they would enable it in the given situation.
In the first scenario the user would lose a functionality if 2FA would not be activated, as it has happened on Steam with their trade and market hold. 
228 of the 462 participants ($49.4\%$) stated that they would very likely activate 2FA ($\mu=5.6$, $\sigma=1.87$, median=6).
The next scenario was the introduction of gameplay advantages, if the user activates 2FA. 
Of all participants, 237 ($51.3\%$) stated that they would use 2FA in this case ($\mu=5.61$, $\sigma=1.93$, median=7).
In the last scenario the users were offered an exclusive visual in-game modification for using 2FA.
Only 181 of the asked people ($39.2\%$) indicated they would very likely start using 2FA to gain said modification ($\mu=5.01$, $\sigma=2.09$).
This stands in contrast to the self-reported reasons for activating 2FA in practice, as reported above.

\subsection{General Population Sample}
\label{sec:res_general}

While we opened the recruitment to people from Canada and the UK, nobody from these countries participated in our survey.
Therefore, we got 313 participants in total, with 155 from the US and 158 from Germany.
After cleaning the data and removing participants who failed the attention check, we retained a total of 288 participants, with 146 being from the US and 142 from Germany.
The complete demographic data can be found in Table~\ref{tab:demographics}.

\begin{table}[t]
    \centering
	\begin{tabular}{lrr}
		 & Gaming Sample & General Sample \\
		 \midrule
		 Total Participants & $462$ & $288$ \\
		 \midrule
		 From the USA & $36\%$ & $51\%$ \\
		 From Germany & $14\%$ & $49\%$ \\
		 From Canada & $7\%$ & $0\%$ \\
		 From the UK &  $7\%$ & $0\%$ \\
		 \midrule
		 Male & $76\%$ & $69\%$ \\
		 Female & $19\%$ & $31\%$ \\
		 Non-binary & $2\%$ & $0\%$ \\
		 No gender data & $2\%$ & $0\%$ \\
		 \midrule
		 Avg. Age & $26.5$ & $32.3$ \\
		 Standard Deviation & $7.2$ & $9.5$ \\
	\end{tabular}
	\caption{Demographic data from both surveys, reported after cleaning the data.}
	\label{tab:demographics}
\end{table}

While $92.7\%$ of participants ($95.2\%$ from US, $90.1\%$ from Germany) stated that they enjoy playing video games, only $62.5\%$ (US: $62.3\%$, DE: $62.7\%$) associated the term \textit{valve} with video games.
For further analysis, we considered the sub-sample who recognized the company Valve as well as enjoyed gaming as our \textit{gaming sub-sample} ($N=179$).
The average age for the gaming sub-sample is at 30 years, while the non-gaming sample average at 35 years.
While the gender ratio is equally divided on all mentioned genders for non-gamers ($53.2\%$ female compared to $46.8\%$ male), there is a much larger imbalance within the gaming sub-sample where $82.1\%$ identified as male and only $17.3\%$ as female.

When asked whether participants use 2FA for at least one of their accounts, $85.7\%$ agreed to do so (US: $85.6\%$, DE: $85.9\%$). 
The adoption rate for people from the gaming sub-sample was $87.7\%$.
A detailed overview of account types and 2FA usage rates can be found in Table~\ref{tab:usage_general}.

\begin{table}[t]
    \centering
    \begin{tabular}{llrrr}
        Service & Overall & 2FA & GSS & 2FA\\
        \midrule
        Online Banking & $94.8\%$ & $65.3\%$ & $96.1\%$ & $67.6\%$\\
        Backup \& Cloud & $73.3\%$ & $14.6\%$ & $76.5\%$ & $17.9\%$\\
        E-Mail & $96.2\%$ & $40.3\%$ & $97.8\%$ & $41.3\%$\\
        Social Media & $88.5\%$ & $23.6\%$ & $88.3\%$ & $23.5\%$\\
        Messaging & $85.1\%$ & $11.8\%$ & $86.6\%$ & $12.8\%$\\
        Online Games & $73.6\%$ & $30.6\%$ & $88.3\%$ & $44.1\%$\\
        Retail & $89.9\%$ & $22.9\%$ & $91.6\%$ & $25.7\%$\\
        Productivity & $63.2\%$ & $8.0\%$ & $68.7\%$ & $9.5\%$\\
        Hosting & $23.6\%$ & $6.3\%$ & $27.9\%$ & $9.5\%$\\
    \end{tabular}
    \caption{Usage and adoption rates for various gaming-related services from the general population sample ($N=288$), and its gaming sub-sample (GSS, $N=179$). 2FA adoption rates relate to the respective total $N$.}
    \label{tab:usage_general}
\end{table}

Overall, we can see that the gaming sub-sample has higher 2FA adoption rates throughout all categories.
Online games have the third highest 2FA adoption rate (after banking and e-mail) for the overall sample, while in the gaming sub-sample, they are placed second highest.

After asking again about selected services, the participants were asked about what 2FA instruments they use.
Again, e-mail and SMS were the most common instruments, with $35.8\%$ of participants using e-mail and $34.4\%$ using SMS at least once a week (gaming sub-sample: $38.0\%$ e-mail, $40.8\%$ SMS).

As for convenience and perceived security, SMS was ranked most convenient ($\mu_{all}=3.216$, $\sigma_{all}=1.783$, $\mu_{GSS}=3.380$, $\sigma_{GSS}=1.697$) and most secure ($\mu_{all}=2.984$, $\sigma_{all}=1.685$, $\mu_{GSS}=2.975$, $\sigma_{GSS}=1.587$).
These ratings are a change from our first survey with a gaming population where Google Authenticator was ranked most convenient, and hardware tokens were ranked most secure (cf. Section~\ref{sec:res_gaming}).

For each Likert scale question we conducted a Mann-Whitney-U test and compared the ratings of the gaming sample with those of the non-gaming sample. The results were corrected using the Bonferroni-Holm method which shows that there are almost no significant differences concerning how both groups rate different 2FA methods. A notable exception to this is the convenience of service specific 2FA apps (such as the Blizzard Authenticator), which was rated significantly higher by gamers (U = 7544, p < 0.001). This implies that gamers are more comfortable or familiar using them.

Regarding reasons for using 2FA, the large majority of $76.0\%$ stated that their primary reason for using 2FA is account security.
This is again in line with the previous study where security was also the major concern for users.

To test which kind of incentive might be interesting for a general population we decided to phrase several ideas for incentives that were inspired by existing incentives, but modified to lose the direct gaming context and used a 5-point Likert scale to ask how likely participants would activate 2FA in this scenario.
While most of these examples were closely connected to the categories we find in gaming-related incentives, i.e. cosmetic enhancements, gameplay advantages, and sanctions, we also added incentives such as one-time payments, discounts, or physical gifts to complement the selection.
Results show that monetary incentives like one-time payments would be most interesting for our participants with an average score of $3.752$ ($\sigma=1.385$, median=4).
When differentiating between the gaming and non-gaming sub-samples, we see that gamers are more interested in gaming-related incentives like gameplay advantages with an average Likert score of $3.38$ ($\sigma=1.48$, median=3) in comparison to $2.67$ ($\sigma=1.47$, median=2) for non-gamers. This is supported by a Mann-Whitney-U test that shows a significant difference between the gaming and non-gaming sub-sample (U = 6568.5, p < 0.001). %p = 0.000001  
There are no other outstanding differences in the ratings for other incentives.
In our first study, participants rated only three gaming-related incentives on a 7-point Likert scale. The results suggested that all were well received by users, although loss of function ($\sigma=5.60$) and gameplay advantages ($\sigma=5.61$) attracted more users than visual modifications ($\sigma=5.01$).

\begin{figure*}[!h]
\centering 
  \subfloat[]{ 
	\includegraphics[width=0.3\textwidth]{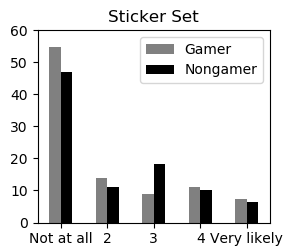}}
     \subfloat[]{ 
\includegraphics[width=0.3\textwidth]{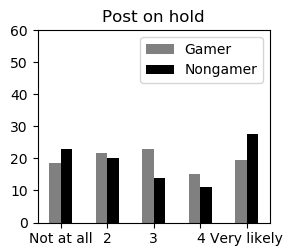}}
\subfloat[]{ 
	\includegraphics[width=0.3\textwidth]{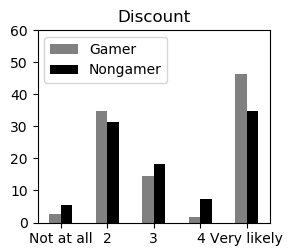}}
    \caption{Answer distributions for the question \textit{How likely is it that you would activate 2FA in the following scenarios?}. Only selected incentives are presented.}
    \label{fig:incentives}
\end{figure*}

A graphical overview on the attractiveness of various incentives is presented in Figure~\ref{fig:incentives}.
We can see a bimodal distribution for the permanent discount scenario with most participants of both groups finding it \textit{very likely} that this kind of incentive would lead them to activate 2FA for an account.
Participants from both groups found sticker sets, which we thought would correspond closely to visual incentives found in online games, very unattractive as an incentive in general.
For the restriction scenario of keeping social media posts on hold for moderation unless a user has activated 2FA, we received very mixed answers.

To see whether incentives increase the adoption rate for 2FA, we looked again at adoption rates by service.
Although most gaming-related accounts we listed in the first survey were replaced by general services, we still asked about Steam and Blizzard accounts regarding 2FA.
While both have mediocre usage rates over the whole population with $54.9\%$ for Steam and only $33.0\%$ for Blizzard, both have high 2FA adoption rates when compared to other services.
Between the gaming and non-gaming sub-samples, we see again that Steam and Blizzard have comparably very high 2FA adoption rates for all specific accounts we asked within the gaming sub-sample, i.e. $22.9\%$ of our gaming participants use 2FA for their Blizzard accounts and $35.2\%$ for Steam while these values in the non-gaming sub-sample are between $3.7\%$ and $5.5\%$.

The 2FA adoption rates for both in the gaming sub-sample are also larger than the respective ones for all participants, where $16.3\%$ stated to have 2FA activated for their Blizzard accounts and $23.3\%$ use 2FA for Steam. 
For both the whole population as well as the gaming sub-sample these values are only topped by online banking, where $45.1\%$  of all participants employ 2FA, and $49.2\%$ of participants within the gaming sub-sample. 
Services such as Paypal, Amazon and Google Mail also have 2FA adoption rates between $28.8\%$ and $36.3\%$ for both groups. 
The other 23 specific services in the list we provided achieved lower adoption rates of at most $16.2\%$.

Overall we performed Fisher's exact tests to compare the general usage rates to the 2FA adoption rates between the gaming and non-gaming samples. While we find no significant differences for specific websites, we find that participants from the gaming group are more likely to adopt 2FA for online game accounts in general (0.342, p = 0.004, Bonferroni-Holm corrected for multiple testing).

\section{Design Space and Concepts for Non-Gaming Incentives}

From the surveys, we learned that account security and account value are seen as large motivators to adopt 2FA.
When it comes to incentive types, we see that both restrictive measures for non-adopters as well as financial or gameplay advantages are rated as convincing for the adoption of 2FA.
This was why we decided to pursue these incentive types further and draft non-gaming examples of these types.
In addition, we also decided to adopt an example for visual modifications, since these were the most common type of incentive seen in the gaming landscape.

The first step in transferring popular incentive types from gaming to non-gaming contexts was reflecting how the design space would change.

Online games have a closed economy and a fixed number of distinct items to acquire.
This nudges players to complete collections of items such as companion pets, and a 2FA incentive can easily hook into this mechanic.
When trying to transfer the incentive of visual modifications or companions, we searched for a similar mechanic that is of only cosmetic (i.e. not functional) value and that comes with a collection or completionist nudge.
We found that stickers in instant messaging are an increasingly popular cosmetic gimmick, and that some people exhibit a similar collection behaviour~\cite{zhou2017goodbye}.
Therefore, we chose to select exclusive messenger stickers as a non-gaming incentive for adopting 2FA.
We thus created a mock-up of a WhatsApp conversation which features two users discussing and presenting an exclusive sticker that was obtained by activating 2FA.

Economic advantages were generally well-accepted by the survey participants (see also Figure~\ref{fig:incentives}); most participants stated that it would be very likely that they activated 2FA if they were either offered a discount in e.g. an online shop or if they received a one-time payment for activation.
In gaming, economic advantages as incentives for activating 2FA in gaming are less frequent and come in different implementations, such as more inventory space (cf. Figure~\ref{fig:examples}).
As we found it hard to model an equivalently ``powerful'' incentive in a non-gaming context,  we chose a discount for an online shop, which is also an economic advantage.
This keeps the design generally applicable as opposed to service-specific economic advantages like extra storage space for a file sharing service.
Therefore, we designed a mock-up of a popular German clothing shop website offering a 5\% discount while 2FA using e-mail as the second factor was activated.

We were also very interested in restriction mechanisms such as those of Steam, so we additionally included a scenario which inhibited platform use without 2FA activated.
While other scenarios were possible, we chose post moderation and restrictions in a social network as an example.
Again, this was so that as many participants as possible would relate to the scenario.
We created two case mock-ups using Facebook as a template.
In the first, the user attempted to post a status update that included another user. 
Since the second person had 2FA activated, but the first user did not, the mock-up does not allow the status to be posted. 
In the second example, the user tried to post something in a public group but was once again stopped and made aware of the fact that before their post was published, a moderator needed to confirm it. 
Both inhibitions could be circumvented by activating 2FA.

\begin{figure*}
    \centering
    \subfloat[Exclusive stickers for a messaging app.]{ 
        \includegraphics[width=0.25\textwidth]{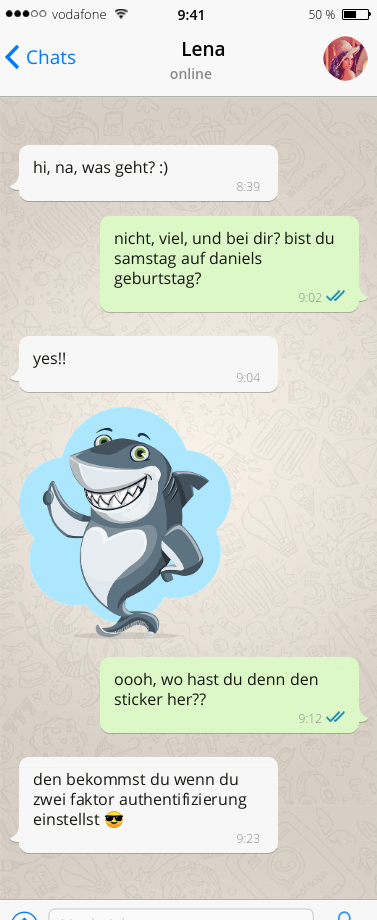}} \quad
    \subfloat[Two social network scenarios: Posts to a group need to be moderated without 2FA (top), tagging a person who uses 2FA is disabled unless the poster themselves enables 2FA, too (bottom). This image was blinded for submission.]{ 
        \includegraphics[width=0.545\textwidth]{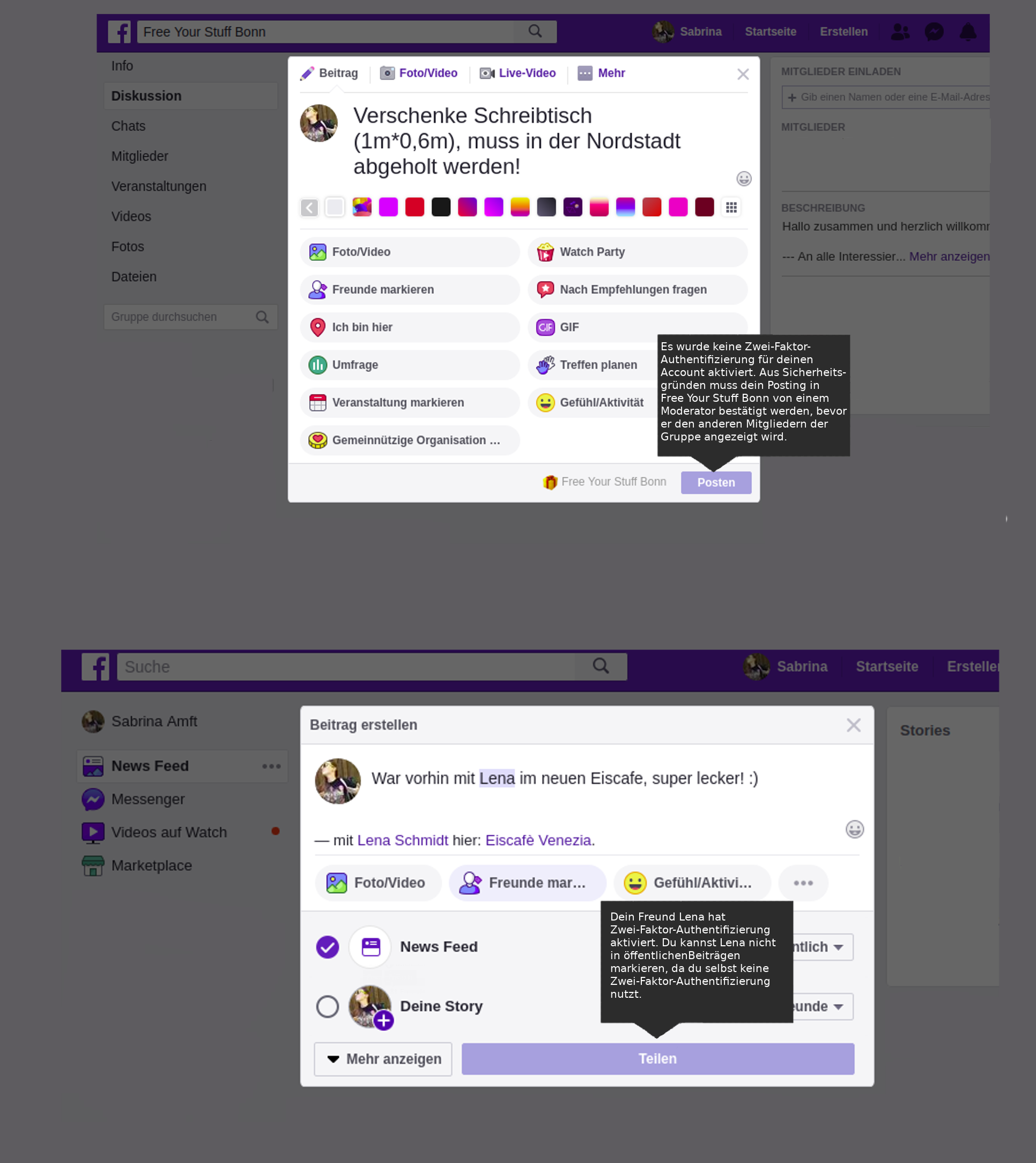}} \quad
    \subfloat[Permanent shop discount while having 2FA activated.]{
        \includegraphics[width=0.8\textwidth]{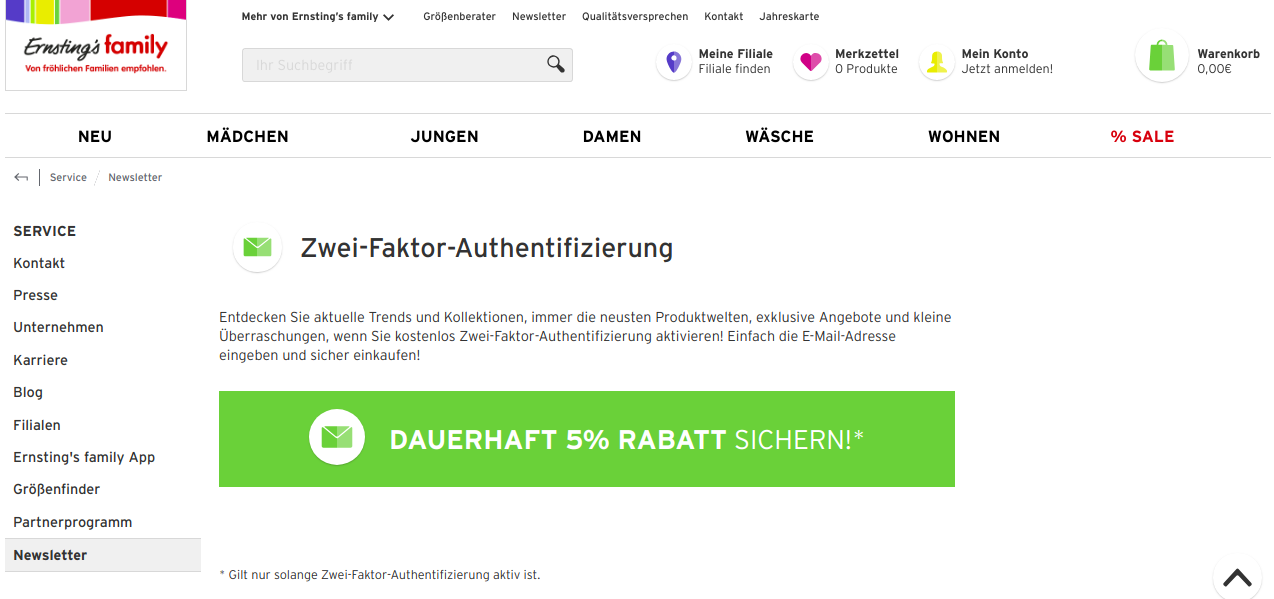}}
    \caption{2FA incentive mock-ups for various services which were designed for the focus group evaluation.} 
    \label{fig:mockups}
\end{figure*}

To summarize, we chose the following incentive types to design artifacts: Economic advantages, inhibited social media, and exclusive sticker sets.
All mock-ups can be found in Figure~\ref{fig:mockups}.

\section{Focus Group Study}

In order to get deeper insights into users' attitudes towards 2FA in general and the various incentives we designed, we decided to conduct a focus group study.
Focus groups are group interviews of 3-6 participants which foster group discussion and are often used for getting impressions on new concepts or ideas~\cite{baxter2015understanding}.
Since our goal was to explicitly evaluate new concepts for 2FA incentives in non-gaming contexts, we chose this method for our evaluation.

\subsection{Methodology}

We tested the mock-up designs (cf. Figure~\ref{fig:mockups}) in a focus group study with 15 participants, who received EUR 10 as compensation for their participation.

The group interview contained discussion of general knowledge and usage experience of various 2FA methods, associated group rating exercises regarding simplicity, ease of use, security and likeliness of adoption.
Afterwards, real 2FA incentives as well as our  mock-ups were presented and discussed.

Three groups of participants were recruited through personal contacts and advertisements placed around the university campus.
All participants were between 18 and 29 years old and from Germany. 
While the first group was an all-male assembly of five computer science students, the second group consisted of three women and two men, who also all studied computer science.
The third group featured three women and two men, all participants were not enrolled into a computer science program.

In the following, we present the most important arguments and findings from our different group sessions.

\subsection{Results}
At the start of each session we asked participants to tell us what they knew about 2FA. 
While both computer science groups were able to name different methods, the third group only knew about SMS and e-mail as well as some examples from banking contexts. 
During the session, they seemed to have different misconceptions about what 2FA was and when it was used as they confused it with e-mails about suspicious account activities or SMS that included account activation codes.

\begin{quote}
    Isn't that sufficient? I mean, you have registered with [your phone number], so it's a kind of two-factor authentication when they send you the confirmation code via SMS and the phone processes this code. (A participant from group 3)
\end{quote}

\begin{figure}
    \centering
    \includegraphics[width=0.35\textwidth]{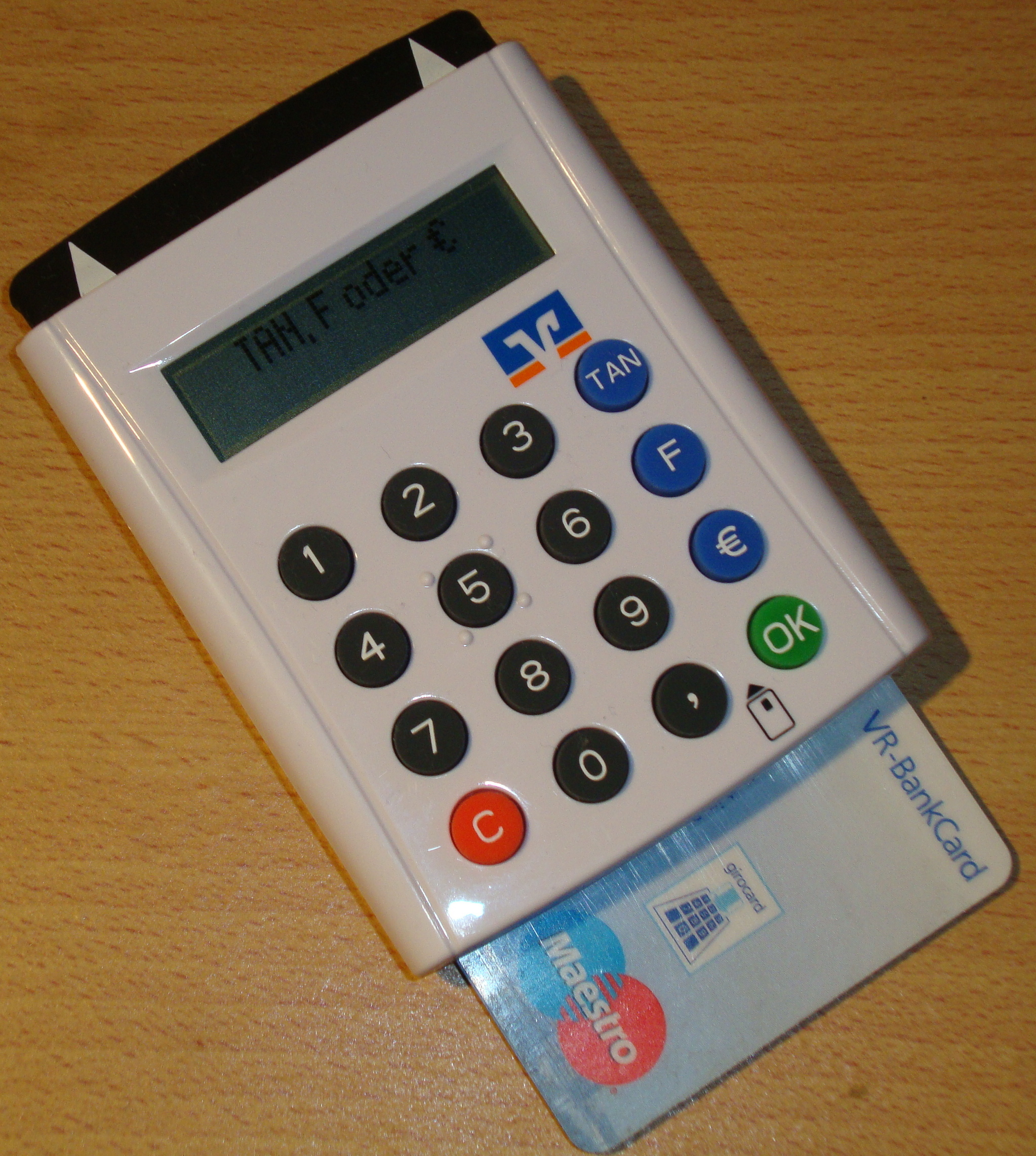} 
    \caption{A \textit{SmartTAN Optic} TAN generator from a German bank. The customer's card is inserted into the device, transaction data is transmitted through optical sensors by holding the token in front of a flickering code on screen. As a fallback mechanism, the data can also be entered through the device's keypad. After the relevant transaction data is displayed on the screen and acknowledged by the user, the device eventually displays the TAN.}
    \label{fig:tan_generator}
\end{figure}

When asked about the properties of different methods, there were major misconceptions about the security of SMS. 
The method was perceived as very secure in all three groups with arguments such as \textit{``since they only arrive on my phone and operate on the SIM card''}, or that \textit{``you cannot read the code on the lock screen, and without my fingerprint nobody can access it''}. 
All three groups also voiced concerns about e-mail being not secure enough as they either thought hacking an e-mail inbox was incredibly easy or that it was purely depending on their password strength. 
This shows a lack of understanding even in the computer science groups that were otherwise able to explain not only the basic concepts of 2FA methods but in some cases also the underlying algorithms. 
Interestingly, although SMS was perceived as more secure than e-mail, most participants seemed to prefer the usage of an e-mail-address for 2FA and did not want to disclose their number. 
Authenticator apps like the Google Authenticator were mostly unknown by our focus group participants.
Some people who identified as gamers reported to use them, for example for their Blizzard accounts.
Hardware tokens were on the other hand very common, as many German banks require token-generated TANs for online banking transactions.
An example of such a token can be seen in Figure~\ref{fig:tan_generator}.
User sentiment about these generators was often negative, as our participants have experienced delays and hindrances in acquiring such a generator from their bank in the past.
In addition, they report problems in generating the TANs by holding the token in front of a flickering code on their computer screen that contains the transaction information.
For example, a participant from group 2 reported \textit{``it's a fifty-fifty chance''} if the device would actually work as intended.
Another participant stated
\begin{quote}
    It depends on what kind of token you have. If I think about my TAN generator, it's horrible. (A participant from group 2)
\end{quote}
While most participants had no previous experience with hardware token, they were perceived as secure. However, participants expressed concerns about this method as the token needs to be carried in person and could easily get lost. Therefore they were rated as less convenient than other methods such as SMS.

In addition, there were often different views on which methods participants would use in different cases.
We saw a clear differentiation between accounts where payment data was attached or shopping history was collected, for these accounts our participants would accept 2FA rather than for ``unimportant'' accounts such as credentials for online forums.
A de-facto consensus was the wish for a selection of 2FA methods, as then every user could cherry-pick the method they liked best.

In general, all groups seemed to agree that while rewards were an interesting approach, especially the sticker set was perceived as not good enough to weigh out the negative sides of 2FA usage, i.e. as the extra effort required and the potential disclosure of contact data. 
They also rejected incentives that were ``too good'': Group 2 and 3 voiced concerns that a company offering rewards for i.e. a phone number required for 2FA might have malicious intents such as selling the number or abusing it for unwanted advertisements.
\begin{quote}
    I would be a little skeptic whether they would sell my [phone number] later on, especially when I get something in return for [enabling 2FA]. (A participant from group 2)
\end{quote}

Participants found the shop discount scenario very appealing, especially since it used e-mail as the second factor instead of SMS or a custom solution.
Participants in group 1 and 2 explained that the shop would have their e-mail address anyway, so enabling 2FA in this case would not come with additional exposure of personal information.
One participant in group 2 argued that this might abuse the situation of lower income households, basically forcing them to adopt 2FA as they would be dependant on the discount.

The inhibition of social media in the Facebook example was partially accepted as a good incentive, although some participants in groups 2 and 3 were unsure about the benefit for the users and stated that they might not use the service at all in this case as they found the inhibition to be annoying.

All three groups stated that websites that either directly handled monetary purposes such as online banking or those that indirectly handled their bank data or valuable items such as Amazon or Steam were likely candidates for 2FA adoption. 
Other personal information was also named, although participants deemed them not as important as websites that dealt with money. 
A participant from group one even explicitly stated that he would not use 2FA for dating apps, although these might hold very intimate information about a user.

Another often mentioned aspect was that participants seemed to weigh the usefulness of 2FA against the potential benefits. 
If a website did not hold important enough information or if a login with 2FA was required too often, they rejected 2FA usage in general. 
However, for i.e. a bank account, users would even use their phone number as this was deemed important enough.

Finally, participants argued that often there would be no need to lure users with rewards, but one for more explanations and educations on what 2FA is and why they should adopt it.

\section{Discussion}

In the following, we will discuss selected topics that emerged from the focus group interview and put them in context with our research question.

\subsection{General Privacy and Security Perception of 2FA}
We found grave misconceptions about the security of SMS. 
SMS messages are sent in plain text and can be easily attacked with low equipment costs~\cite{toorani2008solutions}, which is why NIST declared them insecure as a second factor in 2016, but reverted the statement in a later update~\cite{dudley2017rollback}.
Despite these security flaws, the participants in our survey as well as the focus group interviews tended to regard SMS as very secure and placed a lot of trust in the medium.

Some participant statements in the survey and the focus groups suggest that there might be a differentiation between primitive mobile phone features that use a SIM card (i.e. calling, SMS), and smartphone features that resemble typical computer applications and behaviour (i.e. apps, messaging, and web browsing) within users' mental models.
While the phone features might be regarded as more secure, the smartphone features come with similar risk perceptions as general internet and PC applications~\cite{krombholz2019if}.
News coverage and vulnerability disclosures could have worked toward this narrative, because it is usually the phone operating system or selected apps that are portrayed as insecure, while there is no such coverage about primitive phone features.

Participants also had clear differentiation between which accounts were worth protecting with 2FA.
A common pattern in our focus groups was that when monetary value or sensitive payment data (e.g. SEPA account details) is connected to an account, it becomes worthy of securing it with 2FA.
In contrast, accounts which are rich with personal and intimate data but not associated with money are not deemed worthy of additional protection by our participants.
This confirms prior research on the monetary value of personal data~\cite{DVDgeld, Cvrcek2006study} 

This conception about data being not as worthy as actual money has been discussed before~\cite{grossklags200725}, and it remains open if we as a professional community should see this as a need for better concepts and communication of those, or as a field where strict consumer protection is necessary to soften the impact of these user mental models.

When it comes to handing over phone numbers, we saw some mix-up between account validation purposes and actual 2FA setup.
While companies usually employ account validation via SMS to restrict automated account creation and increase the advertising value of their users' data, 2FA via SMS has the only purpose of making the account more secure.
We regard this tendency as very alarming, since we hypothesize that the practices for user data harvesting have the potential to negatively influence users' mental models of 2FA's security benefits.

\subsection{Incentives in Gaming}
While the adoption rate for 2FA in gaming services that offer incentives are rather high (cf. Table~\ref{tab:usage_gaming}), we see only very small approval when asking directly about the influence of an incentive on adopting 2FA.

It could be that these gaming accounts in question are first and foremost seen as valuable accounts by our participants, since especially Massively Multiplayer Online game (MMO) accounts often contain hundreds of hours of playtime and an assortment of valuable items.
The side-economy of buying and selling actual accounts for often several hundred Dollars on marketplaces like eBay or special platforms like g2g~\cite{g2g} gives this perception additional weight. 
This might make the value of gaming accounts more visible than for example, social media accounts~\cite{parizi2019security}.

Another approach to explain this discrepancy could be increased advertising of 2FA through the measure of incentives.
Usually, gaming publishers release accompanying news and social media posts when introducing an incentive for 2FA adoption~\cite{wownews, fortnitenews, rbsixnews}, this generates publicity and might introduce players to the concept of 2FA who haven't been in contact with this security measure before, thus raising awareness for account security in general. 
These news are often further distributed by major gaming news websites such as Kotaku~\cite{gw2news, gtanews}.
Our findings from the focus group interviews where participants generally wished for more and better education on the subject of 2FA resound with this observation.

Besides social media posts and press releases, visual incentives in an online game are also advertising in itself.
Players see new items, skins, or companions in-game and might start asking around or researching how to acquire them.
This way, they also eventually reach the information about account security and might become aware of the security benefits of adopting 2FA.
However, our focus group interviews have clearly shown that this mechanism only works in closed economies with a focus on collecting rare and different items.
In the general online world, the attractiveness of a sticker set incentive was made highly dependant on the target group, e.g. teenagers.

\subsection{Influence of Incentives on Security}
When an incentive allows for reliable distinction between users who have activated 2FA and those who have not, the service provider might endanger their users who have 2FA not enabled.
A well-designed incentive thus must not allow to filter users by 2FA activation.

Visual incentives in games like skins, emotes, or mini pets can be disabled, switched out, or simply not used by players.
Access to restricted vendors is not visible to outsiders, the same goes for inventory space and in-game wallet.

Steam's trading hold time for 2FA-disabled accounts is visible to trading partners, but initiating a trade needs confirmation from both participating users.

When designing incentives for non-gaming contexts, designers have to keep this security constraint in mind, especially for social features such as moderated posting.

\subsection{Transferability between Gaming and Non-Gaming Contexts}
Video games and especially MMOs have fixed and clear rules of \textit{value}~\cite{lehdonvirta2009virtual}.
Collecting a wide variety of items, especially rare items, is deemed an important meta-goal in online multiplayer games.
This pressure to collect is not as prominent in daily life in Western societies as it is in gaming communities, which has to be taken into account when discussing the transferability of incentives for 2FA.

Our focus group results have shown that especially visual incentives work better in closed-economy contexts than in the daily online world, as sticker sets were dismissed as rather special or only suitable for a narrow audience like teenagers.
The general controversy about incentive types we saw in the group discussions suggests that maybe offering a range of incentives from which participants could pick the one they like best would be a golden way for non-gaming contexts.
However, this would come with increased setup costs for service providers. 

Our focus group results also indicate that there is a certain sweet spot about the power of an incentive.
Participants acknowledged the increased pressure on low-income customers in the context of discount incentives.
While all participants agreed that 2FA was a good thing in general, they were rather torn on incentives that de-facto pressured users into enabling 2FA because of powerful advantages. 
In addition, we observed concerns about unfair advantages or even incentives as bait for data abuse.
When an incentive looked too good to be true, our participants turned skeptical and suspected a bait offer to gather phone numbers or the like.
Regarding the transfer of 2FA incentives into non-gaming contexts, incentives that offer discounts or in-ecosystem advantages such as more space in the cloud might work best.

\subsection{Suggested Incentives for the non-gaming Context}
Gamers have an overall higher adoption rate throughout all categories of services we asked about.
This could indicate that successfully adopting 2FA in at least one field lowers the bar for adoption in other fields.

In regard to our proposed examples for incentives mechanisms in non-gaming contexts (cf. Figure~\ref{fig:mockups}), we found that the shop discount was the most attractive example for our focus group participants.
The monetary discount is a strong pull, and e-mail as a second factor is perceived as non-intrusive.
No additional user data disclosure is needed for setting up 2FA in this case, which was an aspect that our participants highlighted as positive.

In contrast, the other scenarios we proposed were dismissed as not very attractive (sticker set), and too restrictive (Facebook posting restrictions).
We were curious about the attractiveness of cosmetic incentives and thought the sticker feature of many modern messengers would be a good equivalent to gaming contexts, but it turned out that stickers have no such collection effect as weapon skins or mini pets in the gaming context might have.

Drawing from these results, we propose the following design recommendations:
\begin{itemize}
    \item The industry should consider monetary benefits as we found that 2FA incentives for non-gaming contexts to be attached with monetary value where applicable and effective.
    \item Services should offer a set of alternatives as second factors to suit users' needs and their willingness to accept privacy trade-offs for enhanced account security.
    \item Also, as we saw the wish for more education regarding 2FA, we strongly propose some educative text or media to accompany a 2FA campaign in non-gaming contexts, as users are usually not as tech-savvy as gaming populations.
\end{itemize}
Suitable education is orthogonally important as it could help with clarifying and correcting divergent mental models about how 2FA works and what privacy risks are associated with it.

\section{Limitations}
As every scientific work, this one is not without its limitations.

First and foremost, all data we collected through our surveys and focus group study was self-reported.
It is known that people try to put themselves in a better light in such cases, especially when reporting about security and privacy practices and motivations.
This phenomenon could have skewed our results, so field study work is needed to confirm (or reject) our findings.
In addition, some users might not know that they already use some kind of 2FA in their daily lives, which would result in a skew in the other direction.

While both surveys we conducted were similar, there were some modifications to the second run. Furthermore, both samples were conducted on different platforms, several months apart and with different motivations as the gaming sample received no compensation whereas the general sample was rewarded with USD 2.00. All of these differences might have had an influence on the answers our participants gave.

Our focus groups only portray a very narrow cultural and demographic sample, since all participants were rather young and from Germany.
Students are known to be more tech-savvy and innovation-friendly in general, and German cultural and societal values like an emphasis on privacy~\cite{krasnova2012self} might have influenced our results.
Further research with more diverse sets of participants is clearly needed and much appreciated.

\section{Conclusion and Future Work}

In this paper, we presented the design, evaluation, and interpretation of three user studies about incentives for adopting two-factor authentication (2FA).
We conducted two surveys with multi-national samples, within a gaming-focused population and a general population sourced by crowdworking.
Based on the results of these surveys, we designed three novel concepts for incentive mechanisms in non-gaming contexts, namely a sticker set reward for a messenger application, the revocation of posting restrictions in certain areas of a social network, and a permanent discount for an online shop as long as the customer has 2FA enabled.

We found that there is no ``one fits all'' solution.
This confirms previous work on authentication~\cite{Weir2009}.
Participants expressed needs to minimize the risk of losing access to a second factor as well as portability of said factor.
They were concerned about disclosing private information like phone numbers for SMS 2FA and favoured a selection of 2FA instruments to choose from, based on their needs and the perceived importance of the respective account they want to protect.

From the three designs we proposed, the online shop discount incentive was considered most attractive, while the sticker set turned out uninteresting for most participants.
This suggests that the apparent effectiveness of cosmetic incentives which can be found in gaming contexts is not applicable per se to non-gaming contexts.
Furthermore, participants favoured a selection of different 2FA mechanisms instead of one ``golden way''.
From these experiences, we formulated three actionable recommendations for deploying incentivized 2FA: Use monetary incentives where applicable, as they have the strongest pull.
Offer alternatives to suit users' individual needs.
Educate about 2FA and its benefits in general.

Future research could attach here and evaluate different combinations of 2FA mechanisms that cover a wide range of audience.
In addition, field studies about the actual adoption likeliness and user behaviour in the wild are needed.

\begin{acknowledgement}
We thank all our survey and focus group participants. Thanks to Matthew Smith for valuable feedback and supervision during the early conception of this work.

%TODO
Portions of the materials used are trademarks and/or copyrighted works of Epic Games, Inc. All rights reserved by Epic. This material is not official and is not endorsed by Epic.

World of Warcraft ©2004 Blizzard Entertainment, Inc. All rights reserved. World of Warcraft, Warcraft and Blizzard Entertainment are trademarks or registered trademarks of Blizzard Entertainment, Inc. in the U.S. and/or other countries.
\end{acknowledgement}

\newpage
\appendix

\section{Gaming Survey}

\textit{Here we present the questionnaire for the first survey that was distributed in online gaming communities.
This survey was designed and conducted as part of a student project.}

The usage of services in the internet is rising. Historically the user accounts of such services are secured by using a password. 
To increase the security, two factor authentication (2FA) is used.
When 2FA is used, the user needs another information to login. 
As an example, this could be a code provided via a
mobile app or a SMS sent to a specified phone number.
In many online games or game services it is feature-wise beneficial for the user to activate 2FA.
The goal of this survey is to gather statistical data how the incentives or restrictions have influenced the 2FA adoption rate within the user base.
This survey should only take you a few minutes. Thanks in advance for participating!

\paragraph*{1. Demographics}
Please fill in your age, gender and your current country of residence. \textit{free text entry}

\paragraph*{2. Which services do you use?}
\begin{itemize}
    \item Blizzard Battle.net
    \item Discord
    \item Facebook
    \item GOG.com
    \item Guild Wars 2
    \item Nintendo Account
    \item Origin
    \item Playstation Network
    \item Reddit
    \item Slack
    \item Steam
    \item Telegram
    \item Twitter
    \item Wargaming
    \item WhatsApp
    \item XBox Live
    \item I don’t use any services on this list
\end{itemize}

\paragraph*{3. For which services do you have 2FA activated?}
Please mark the services you are actively using 2FA for.
\begin{itemize}
    \item Blizzard Battle.net
    \item Discord
    \item Facebook
    \item GOG.com
    \item Guild Wars 2
    \item Nintendo Account
    \item Origin
    \item Playstation Network
    \item Reddit
    \item Slack
    \item Steam
    \item Telegram
    \item Twitter
    \item Wargaming
    \item WhatsApp
    \item XBox Live
    \item Other (free text entry)
    \item I do not have 2FA activated
\end{itemize}

\paragraph*{4. Which methods of 2FA do you use?}
\begin{itemize}
    \item SMS
    \item Email
    \item Google Authenticator
    \item Specific app solution, e.g. Blizzard Authenticator
    \item Hardware-Token
    \item Other
    \item I don’t use any 2FA methods
\end{itemize}

\paragraph*{5. How would you rate the following methods of 2FA regarding their convenience?}
(7-point Likert scale from \textit{not convenient} to \textit{very conventient} with \textit{don't know} option)
\begin{itemize}
    \item SMS
    \item Email
    \item Google Authenticator
    \item Specific app solutions, e.g. Blizzard Authenticator
    \item Hardware-Token
\end{itemize}

\paragraph*{6. How would you rate the following methods of 2FA regarding their security?}
(7-point Likert scale from \textit{not secure} to \textit{very secure} with \textit{don't know} option)
\begin{itemize}
    \item SMS
    \item Email
    \item Google Authenticator
    \item Specific app solutions, e.g. Blizzard Authenticator
    \item Hardware-Token
\end{itemize}

\paragraph*{7. Why did you activate 2FA?}
Please mark your primary reasons why you have activated 2FA.
\begin{itemize}
    \item Account security
    \item High monetary value is attached to the account
    \item Gameplay advantage, e.g. an exclusive in-game shop
    \item Visual bonus, e.g. an exclusive in-game pet
    \item To circumvent a restriction, e.g. Steams Community Market Trading hold
    \item Other (free text entry)
    \item I do not have 2FA activated
\end{itemize}

\paragraph*{8. If an incentive convinced you to activate 2FA, ...} ... how easy was the activation of 2FA? (7-point Likert scale from \textit{very hard} to \textit{very easy} with \textit{I wasn't convinced} option)

\paragraph*{9. If an incentive convinced you to activate 2FA, ...} ... how convenient is the usage of 2FA?
(7-point Likert scale from \textit{not convenient} to \textit{very convenient} with \textit{I wasn't convinced} option)

\paragraph*{10. How likely is it that you would activate 2FA in the following scenarios?}
(7-point Likert scale from \textit{not likely} to \textit{very likely})
\begin{itemize}
    \item You would lose a previously available feature for not activating 2FA (e.g. When Steams Community Market Trading hold was introduced)
    \item You could gain a gameplay advantage for using 2FA (e.g. additional inventory slots, exclusive shop)
    \item You could gain an exclusive visual modification for using 2FA (e.g. companion, special skin)
\end{itemize}

\paragraph*{11. What is the probability of you deactivating 2FA, if you could keep the gained benefit(s)?}
(7-point Likert scale from \textit{not likely} to \textit{very likely})
Probability of you deactivating 2FA

\paragraph*{12. Why would you deactivate 2FA?}
Please mark all applicable answers
\begin{itemize}
    \item I don’t like the additional steps required to login
    \item I don’t care about the additional security layer
    \item I think the account is safe enough without 2FA
    \item Other (free text entry)
    \item I would not deactivate it
\end{itemize}

\section{General Population Survey}
\textit{Here we present the questionnaire for the second survey that was distributed on Amazon MTurk.
This survey was designed and conducted as part of a Master's thesis.}

This survey will ask you questions about your online behavior and different security mechanisms as part of a research
project at University of Bonn. The results will be used to research and improve existing security mechanisms. Please
read the questions carefully and answer honestly. We estimate this will take you 10-15 minutes.\\
By completing this survey you consent to the collection and evaluation of your answers. This will only be shared as part
of our project and only with researchers of University of Bonn. The published results will be anonymized. Leaving the
survey without finishing it equals withdrawing your consent, although you can return to finish it as long as the project is
not completed.\\
If you have any questions or feedback regarding this survey please feel free to contact Sabrina Amft, Karoline Busse
or Emanuel von Zezschwitz.\\
Thank you for participating!

\paragraph*{1. Age}
Please fill in your age. \textit{natural number entry}

\paragraph*{2. Demographics}
Please fill in your gender and your current country of residence. \textit{free text entry}

\paragraph*{3. What do understand under the term 'Valve'?}
\begin{itemize}
    \item A metal piece used to block or release a pipe
    \item The company behind a well-known game shop and different video games
    \item A TV Show about metal works
    \item A brand for summer clothing
    \item None of the above
\end{itemize}

\paragraph*{4. Do you enjoy playing video games?} \textit{single selection}
\begin{itemize}
    \item Yes
    \item No
\end{itemize}

\paragraph*{5. Which kinds of online services do you make use of?}
\begin{itemize}
    \item Online-Banking
    \item Backups and Clouds (e.g. Dropbox, Google Drive)
    \item E-Mail (e.g. Gmail)
    \item Social Media (e.g. Facebook, Twitter)
    \item Messaging (e.g. Skype, Whatsapp)
    \item Games (e.g. Battle.net)
    \item Retail (e.g. Amazon, eBay)
    \item Productivity (e.g. Google Docs)
    \item Hosting-Services (e.g. Amazon Web Services)
    \item Other (please specify) 
\end{itemize}

\paragraph*{6. Please mark the services you actively use.}
\begin{itemize}
    \item Amazon
    \item Blizzard Battle.net
    \item Discord
    \item Dropbox
    \item eBay
    \item Facebook
    \item Google Drive
    \item Google Mail
    \item Instagram
    \item Kickstarter
    \item LinkedIn
    \item Nintendo Account
    \item OneDrive
    \item Online-Banking
    \item Origin
    \item Patreon
    \item Paypal
    \item Playstation Network
    \item Reddit
    \item Signal
    \item Skype
    \item Steam
    \item Telegram
    \item Twitch
    \item Twitter
    \item WhatsApp
    \item XBox Live
    \item Yahoo Mail
    \item Youtube
    \item I don’t use any services on this list
    \item Other (please specify) 
\end{itemize}

\paragraph*{Two-factor authentication (2FA)} is a security mechanism that requires a second piece
of information (a second factor) if someone tries to log into an account. This is used to
increase confidence that the person requesting access is really you. Such information is
often a single-use code that is communicated via e.g. SMS, e-mail or apps such as
Google Authenticator.

\paragraph*{7. Do you use two-factor authentication for any of your online accounts?} \textit{single selection, participants who answered with No were forwarded to question 15.}
\begin{itemize}
    \item Yes
    \item No
    \item I don't know
\end{itemize}

\paragraph*{8. Do you use two factor authentication (2FA) for the following?}
\begin{itemize}
    \item Online-Banking
    \item Backups and Clouds (e.g. Dropbox, Google Drive)
    \item E-Mail (e.g. Gmail)
    \item Social Media (e.g. Facebook, Twitter)
    \item Messaging (e.g. Skype, Whatsapp)
    \item Games (e.g. Battle.net)
    \item Retail (e.g. Amazon, eBay)
    \item Productivity (e.g. Google Docs)
    \item Hosting-Services (e.g. Amazon Web Services)
    \item Other (free text answers from question 5)
    \item Other (please specify) 
\end{itemize}

\paragraph*{9. Please mark the services you are actively using 2FA for} \textit{Answers are carried over from question 6.}
    \begin{itemize}
    \item Amazon
    \item Blizzard Battle.net
    \item Discord
    \item Dropbox
    \item eBay
    \item Facebook
    \item Google Drive
    \item Google Mail
    \item Instagram
    \item Kickstarter
    \item LinkedIn
    \item Nintendo Account
    \item OneDrive
    \item Online-Banking
    \item Origin
    \item Patreon
    \item Paypal
    \item Playstation Network
    \item Reddit
    \item Signal
    \item Skype
    \item Steam
    \item Telegram
    \item Twitch
    \item Twitter
    \item WhatsApp
    \item XBox Live
    \item Yahoo Mail
    \item Youtube
    \item I don’t use any services on this list
    \item Other (free text answers from question 6)
    \item Other (please specify) 
\end{itemize}

\paragraph*{10. How often do you use the following 2FA methods?} (5-point Likert scale with the options \textit{Daily}, \textit{Weekly}, \textit{Monthly}, \textit{Sometimes}, \textit{I don't use any 2FA mechanisms})
\begin{itemize}
    \item SMS
    \item E-Mail
    \item Google Authenticator
    \item Specific app solutions, e.g. Blizzard Authenticator
    \item Hardware token e.g. a SmartCard or Yubikey
\end{itemize}

\paragraph*{11. How would you rate the following methods of 2FA regarding their convenience?}
(5-point Likert scale from \textit{very convenient} to \textit{not convenient} with \textit{I don't know} option)
\begin{itemize}
    \item SMS
    \item E-Mail
    \item Google Authenticator
    \item Specific app solutions, e.g. Blizzard Authenticator
    \item Hardware token e.g. a SmartCard or Yubikey
\end{itemize}

\paragraph*{12. If you perceived the convenience of one or more methods to be low, please tell us why.} \textit{free text entry}

\paragraph*{13. How would you rate the following methods of 2FA regarding their security?}
(5-point Likert scale from \textit{very secure} to \textit{not secure} with \textit{I don't know} option)
\begin{itemize}
    \item SMS
    \item E-Mail
    \item Google Authenticator
    \item Specific app solutions, e.g. Blizzard Authenticator
    \item Hardware token e.g. a SmartCard or Yubikey
\end{itemize}

\paragraph*{14. If you perceived the security of one or more methods to be low, please tell us why.} \textit{free text entry}

\paragraph*{15. What is the primary reason why you would activate 2FA?} \textit{single selection}
\begin{itemize}
    \item Account security
    \item High monetary value is attached to the account
    \item Functional advantage, e.g. new or enhanced features
    \item Visual bonus, e.g. a sticker set or an exclusive in-game pet
    \item To circumvent a restriction
    \item I don't use 2FA
    \item Other (please specify)
\end{itemize}

\paragraph*{16. If an incentive convinced you to activate 2FA, \ldots}
... how easy was the activation of 2FA? (5-point Likert scale from \textit{very} to \textit{not at all} with \textit{I don't know} option)

... how convenient is the usage of 2FA? (5-point Likert scale from \textit{very} to \textit{not at all} with \textit{I don't know} option)

\paragraph*{17. How likely is it that you would activate 2FA in the following scenarios?} (5-point Likert scale from \textit{very likely} to \textit{not likely} with \textit{I don't know} option)\textit{first part of the question, item order was randomized}

\begin{itemize}
    \item You could gain a functional advantage for using 2FA (e.g. new features, exclusive shop)
    \item You are offered a sticker set for your favorite social media/messenger
    \item Posts including media (e.g. pictures) are kept on hold until reviewed by a moderator if 2FA is not activated
    \item You receive a small physical gift as a thank-you (e.g. keychain of your choice)
    \item You would lose a previously available feature (e.g. posting status updates, exclusive sales)
    \item You could gain a gameplay advantage for video games such as World of Warcraft (e.g. additional inventory slots, exclusive shop)
\end{itemize}

\paragraph*{18. How likely is it that you would activate 2FA in the following scenarios?} (5-point Likert scale from \textit{very likely} to \textit{not likely} with \textit{I don't know} option) \textit{second part of the question, item order was randomized}

\begin{itemize}
    \item Please choose "very likely" for this question to let us know you're still paying attention.
    \item You could gain an exclusive visual modification for using 2FA (e.g. a pet or costume for video game characters)
    \item You are offered a special offer (e.g. small permanent discount)
    \item It is not possible to access or interact with certain profiles if you do not activate 2FA
    \item You receive a one time bonus payment for using 2FA.
\end{itemize}

\paragraph*{19. What is the probability of you deactivating 2FA, if you could keep the gained benefit(s) from activating it?}
(5-point Likert scale from \textit{very high} to \textit{very low})

\paragraph*{20. What is the probability of you keeping 2FA active if otherwise you would loose the gained benefit(s) from activating it?}
(5-point Likert scale from \textit{very high} to \textit{very low})

\paragraph*{21. Why would you deactivate 2FA?}
\begin{itemize}
    \item I don’t like the additional steps required to login
    \item I don't think that 2FA will help increase the security of my account
    \item I don't need additional security mechanisms
    \item I would not deactivate it
    \item It's not working properly for me (e.g. delays with code delivery)
    \item Other (please specify)
\end{itemize}

\paragraph*{22. Do you have any further remarks about this study or its topic?} \textit{free text entry}

\paragraph*{23. Please enter the following code in Amazon MTurk to help us verify that you completed the survey} \textit{a code was shown}
Yes, I copied the code to MTurk.

Thank you for completing this survey! 
Your answers were transmitted, you may close the browser window or tab now.
\bibliographystyle{SIGCHI-Reference-Format}
\bibliography{bib}
\end{document}